\pdfoutput=1

\documentclass[11pt,twoside,a4paper,cmspaper,final,collab]{cms-tdr}
\def\svnVersion{365580}\def\cmsCernNoTag{CERN-EP/2016-039}\def\cmsCernDate{\today}\def\cmsMessage{Published in Physics Letters B as \href{http://dx.doi.org/10.1016/j.physletb.2016.05.087}{\doi{10.1016/j.physletb.2016.05.087.}}}

\begin{document}\cmsNoteHeader{HIG-15-001}

\hyphenation{had-ron-i-za-tion}
\hyphenation{cal-or-i-me-ter}
\hyphenation{de-vices}
\RCS$Revision: 342988 $
\RCS$HeadURL: svn+ssh://svn.cern.ch/reps/tdr2/papers/HIG-15-001/trunk/HIG-15-001.tex $
\RCS$Id: HIG-15-001.tex 342988 2016-05-15 13:06:34Z alverson $
\newlength\cmsFigWidth
\ifthenelse{\boolean{cms@external}}{\setlength\cmsFigWidth{0.85\columnwidth}}{\setlength\cmsFigWidth{0.4\textwidth}}
\ifthenelse{\boolean{cms@external}}{\providecommand{\cmsLeft}{top\xspace}}{\providecommand{\cmsLeft}{left\xspace}}
\ifthenelse{\boolean{cms@external}}{\providecommand{\cmsRight}{bottom\xspace}}{\providecommand{\cmsRight}{right\xspace}}
\providecommand{\tauh}{\ensuremath{\PGt_\mathrm{h}}\xspace}

\cmsNoteHeader{HIG-15-001}
\title{Search for neutral resonances decaying into a Z boson and a pair of b jets or $\tau$ leptons}

\author{The CMS Collaboration}

\date{\today}

\abstract{A search is performed for a new resonance decaying into a lighter resonance and a Z boson. Two channels are studied, targeting the decay of the lighter resonance into either a pair of oppositely charged $\tau$ leptons or a \bbbar pair. The Z boson is identified via its decays to electrons or muons. The search exploits data collected by the CMS experiment at a centre-of-mass energy of 8\TeV, corresponding to an integrated luminosity of 19.8\fbinv. No significant deviations are observed from the standard model expectation and limits are set on production cross sections and parameters of two-Higgs-doublet models.}

\hypersetup{%
pdfauthor={CMS Collaboration},%
pdftitle={Search for neutral resonances decaying into a Z boson and a pair of b jets or tau leptons},%
pdfsubject={CMS},%
pdfkeywords={CMS, physics, Higgs, 2HDM, BSM, b-tagging, tau, lepton}}

\maketitle 

\section{Introduction}\label{sec:introduction}

The observation of a new particle with a mass of approximately 125\GeV was
reported by the ATLAS and CMS experiments
at the CERN LHC in the WW, ZZ and $\gamma\gamma$ final
states~\cite{Aad:2012tfa,Chatrchyan:2012ufa,Chatrchyan:2013lba}.
Evidence of the decay of the particle to pairs of fermions ($\tau\tau$
and $\bbbar$) has also been reported in Refs.~\cite{Chatrchyan:2014vua,Aad:2014xzb,Aad:2015vsa}.
The measurements of branching fractions, production rates, spin and parity are all consistent with the predictions for the standard model (SM) Higgs boson~\cite{Khachatryan:2014jba,Aad:2015gba},
wherein a single doublet of Higgs fields is present. However, additional Higgs bosons are expected in simple extensions of the SM scalar sector, such as models with two Higgs-boson
doublets (2HDMs)~\cite{Branco:2011iw}. These models predict five physical Higgs particles that arise as a consequence of the electroweak symmetry-breaking mechanism:
two neutral CP-even scalars (h, H), one neutral CP-odd pseudoscalar (A), and two charged scalars (\Hpm).

An important motivation for 2HDMs is that such models provide a way to accommodate the asymmetry between matter
and antimatter observed in the universe~\cite{Branco:2011iw,Dorsch:2014qja}.
An extension of the SM scalar sector with two Higgs boson doublets would also naturally arise in supersymmetry~\cite{Martin:1997ns,Bagnaschi:2039911},
which requires a scalar structure more complex than a single doublet.
Axion models~\cite{Kim:1986ax} provide a strong interaction that does not violate CP symmetry
and give rise to an effective low-energy theory with two Higgs doublets. Finally, it has recently been noted~\cite{Broggio:2014mna} that certain
realisations of 2HDMs can accommodate the muon g--2 anomaly~\cite{Jegerlehner:2009ry} without violating present theoretical and experimental constraints.

In the most general case, 14 parameters describe the scalar sector of a 2HDM~\cite{Branco:2011iw}.
Only six free parameters remain once the experimental observations are included
by imposing the so-called $\mathbb{Z}_2$ symmetry to suppress flavour changing neutral currents,
and by fixing both the values of the mass of the recently discovered SM-like Higgs boson (125\GeV)~\cite{Aad:2015zhl} and the electroweak vacuum expectation value (246\GeV).
The compatibility of a SM-like Higgs boson with 2HDMs is possible in the so-called alignment limit.
The alignment limit is reached when $\cos(\beta-\alpha)\to0$, where \tanb is the ratio of the vacuum expectation values
and $\alpha$ is the mixing angle of the two Higgs doublets.
In such a regime, one of the CP-even scalars, h or H, is identified with the SM-like Higgs boson.
A recent theoretical study~\cite{Dorsch:2014qja} has shown that, in this limit,
a large mass splitting ($>$100\GeV) between the A and H bosons would favour the electroweak phase transition that would be at the origin of baryogenesis in the early universe,
satisfying thereby the currently observed matter-antimatter asymmetry.
In this context, the most frequent decay mode of the pseudoscalar A boson would be $\PSA\to\Z\PH$.
Since the analysis strategy presented in this paper is independent of the assumed model and parity of the resonance,
the results can also be interpreted in the reversed topology $\PH\to\Z\PSA$, where the expected 2HDM mass hierarchy
is inverted and the mass of A is expected to be light~\cite{Gerard:2007kn}. For both topologies, the lighter
scalar resonance (A or H) is not identified with the SM-like Higgs boson.

This paper describes the first CMS search for a new resonance decaying into a lighter resonance and a Z boson.
Two searches are performed, targeting the decay of the lighter resonance into either a pair of oppositely charged $\tau$ leptons
or a \bbbar pair. In both cases, the Z boson is identified via
its decay into a pair of oppositely charged electrons or muons (light leptons), labelled in the text by the symbol $\ell$.
The choice of \bbbar and $\PGt\PGt$ final states is motivated
by the large branching fractions predicted in most of the 2HDM phase
space~\cite{Grinstein:2013npa}. For the $\ell\ell\PGt\PGt$ channel,
the following $\tau\tau$ final states are considered: \Pe\Pgm,
$\Pe\tauh$, \Pgm\tauh, and \tauh\tauh, where
$\tauh$ indicates the decays $\tau \to \text{hadrons}+\nu_{\tau}$.
Given its sensitivity to the 2HDM parameter space region where
$\cos(\beta-\alpha)\approx0$, the search presented in this paper is
complementary to other related searches performed in the same final
state by the ATLAS and CMS collaborations~\cite{HIG14-011,Aad:2015wra}.

\section{The CMS detector}\label{sec:cms}

The central feature of the CMS apparatus is a superconducting solenoid
of 6\unit{m} internal diameter, providing a magnetic field of 3.8\unit{T}.
Located in concentric layers within the solenoid volume are a silicon pixel and strip tracker,
a lead tungstate crystal electromagnetic calorimeter (ECAL),
and a brass and scintillator hadron calorimeter (HCAL),
each composed of one barrel and two endcap sections.
These layers provide coverage up to a pseudorapidity $\abs{\eta}=2.5$.
Extensive forward calorimetry complements are provided by the endcap detectors for $\abs{\eta}<5.2$.
Combining the energy measurement in the ECAL with the measurement in the tracker, the momentum resolution for electrons
with $\pt\approx45\GeV$ from $\Z\to\Pe\Pe$ decays ranges from 1.7\% for nonshowering electrons in the barrel
region to 4.5\% for showering electrons in the endcaps~\cite{Khachatryan:2015hwa}.
Muons are measured in gas-ionisation detectors embedded in the steel flux-return yoke outside the solenoid.
They cover the pseudorapidity range $\abs{\eta}<2.4$, with detection planes made using three technologies: drift tubes,
cathode strip chambers, and resistive plate chambers. Matching muons to tracks measured in the silicon tracker results in a
relative transverse momentum resolution for muons with $20<\pt<100\GeV$ of 1.3--2.0\% in the barrel and better than 6\% in
the endcaps~\cite{CMS-PAPER-MUO-10-004}.
The first level of the CMS trigger system uses information from the calorimeters and muon detectors
to select the most interesting events. A high-level trigger processor farm decreases
the event rate from approximately 100\unit{kHz} to 600\unit{Hz} before data storage.
A more detailed description of the CMS detector, together with a definition of the coordinate system and kinematic variables,
can be found in Ref.~\cite{CMS:2008zzk}.
\section{Data and simulated samples}\label{sec:sample}

The data used for this search were collected by the CMS experiment at $\sqrt{s}= 8\TeV$, and correspond to a
total integrated luminosity of 19.8\fbinv.
The average number of interactions per bunch crossing (pileup) in the data was 21~\cite{Chatrchyan:2014mua}.
Events were selected using dielectron and dimuon triggers~\cite{Khachatryan:2015hwa,CMS-PAPER-MUO-10-004}.
These triggers have \pt thresholds of 17 and 8\GeV for the leading and subleading lepton respectively, and require relatively loose reconstruction and identification criteria.

The main SM background processes giving rise to prompt leptons are W/Z+jets, \ttbar\!+jets, tW, and diboson production (WW, ZZ, and $\PW\PZ$).
The SM background contribution from ZZ is generated at next-to-leading order (NLO) with \POWHEG~1.0~\cite{Alioli:2010xd} for the $\ell\ell\PGt\PGt$~channel and using the leading-order (LO) \MADGRAPH~5.1 Monte Carlo (MC) program~\cite{MADGRAPH}, matched to \PYTHIA~6.4~\cite{Sjostrand:2006za} for
the parton showering and hadronization, for the $\ell\ell\PQb\PQb$ channel. Single top quark events are generated at NLO using \POWHEG~1.0.
Simulated events for other samples are obtained using the \MADGRAPH~5.1 MC matched to \PYTHIA~6.4. The \PYTHIA parameters affecting the description
of the underlying event are set to those of the Z2* tune~\cite{Chatrchyan:2011id}.
All generators used for processes including $\PGt$ leptons in the final state are interfaced with \TAUOLA~2.4~\cite{Jadach:1993hs} for the simulation of the $\PGt$ decays.
The detector response is simulated using a detailed description of CMS, based on the \textsc{Geant4} toolkit~\cite{Agostinelli:2002hh}.
The simulated samples account for contributions from pileup collisions that reflect the distributions observed in data.
The trigger and reconstruction efficiency in the simulation is rescaled by as much as 2\% in order to match that measured in the data~\cite{Chatrchyan:2014mua}.

Two benchmark 2HDM processes are considered as signal: $\PH\to\Z\PSA$ and $\PSA\to\Z\PH$, where the lightest boson (pseudoscalar or scalar, according to the process) can decay to $\tau\tau$ or \bbbar, and the Z decays to $\ell\ell$. The \MADGRAPH~5.1 program, interfaced to \PYTHIA~6.4 and \TAUOLA~2.4, was used to generate signal samples corresponding to different values of A and H masses ($m_{\PSA}$ and $m_{\PH}$, respectively).
The same properties of the SM Higgs boson are assigned to the lightest scalar boson, h, and its mass $m_{\Ph}$ is fixed at 125\GeV.
The identification of the observed Higgs boson, together with all its measured properties, with the scalar $\Ph$ constrains the
phenomenologically reliable parameter space regions to not depart from the SM-like condition $\cos(\beta-\alpha)\approx0$. This corresponds to the
so-called alignment limit~\cite{Gunion:2002zf}.
Considering the parameter space still allowed by direct searches~\cite{Bagnaschi:2039911}, the chosen values
for $\cos(\beta-\alpha)$ and \tanb are 0.01 and 1.5, respectively, and type-II Yukawa couplings are assumed for the benchmark processes.

The masses of the charged Higgs bosons ($m_{\PH^{\pm}}$) are kept equal to the
highest mass involved in the signal process ($m_{\PH}$ or $m_{\PSA}$) to preserve
the degeneracy $m_{\PH^{\pm}}^{2} \approx m_{\mathrm{\PH/\PSA}}^{2}$~\cite{Gerard:2007kn}, denoting with $m_{\mathrm{\PH/\PSA}}$ the mass of the scalar $\PH$ or the mass of the pseudoscalar $\PSA$. The value of the $m_{12}$ parameter, the soft $\mathbb{Z}_2$ symmetry breaking mass, was set to $m_{12}^2=m_{\PH^{\pm}}^2 \,
\tanb/(1+\tan^2\beta)$, according to the minimal supersymmetric standard model (MSSM) parametrisation~\cite{Martin:1997ns}. The value of the complex couplings $\lambda_{6}$ and $\lambda_{7}$ in this parametrisation are set to zero, in order to avoid tree-level CP violation.
The production cross sections, used for the normalisation of the signal samples, are computed using the \textsc{SusHi}~1.4
program~\cite{Harlander:2012pb}, which provides next-to-next-to-leading-order (NNLO) predictions.
The branching fraction for the heavy and light Higgs bosons are
obtained using the \textsc{2hdmc}~1.6 program~\cite{Eriksson:2009ws}, following the
guidelines in Refs.~\cite{harlander:HigXSECWG, Hespel:2015zea}.

The signal benchmark where the light boson decays into
$\PGt\PGt$ is simulated for values of $m_{\mathrm{\PH/\PSA}}$ and $m_{\mathrm{\PSA/\PH}}$ varying in the ranges 200--1000 and 15--900\GeV, respectively, with
the constraint $m_{\mathrm{\PH/\PSA}}>m_{\mathrm{\PSA/\PH}} + m_{\Z}$.
For the $\ell\ell\PQb\PQb$ analysis the lower bound for the invariant mass $m_{\mathrm{\PSA/\PH}}$ goes down to 10\GeV.
The region where $m_{\PH}$ is smaller than $m_{\Ph}$ is not pertinent in this
model.

\section{Event reconstruction and selection}\label{sec:evtReconstruction}

Event reconstruction is based on the particle-flow algorithm~\cite{CMS-PAS-PFT-09-001,CMS-PAS-PFT-10-001},
which exploits information from all the CMS subdetectors to identify and reconstruct individual
particles in the event: muons, electrons, photons, charged and
neutral hadrons. Such particles are algorithmically combined to form the jets,
the $\tauh$ candidates, the missing transverse momentum \ptvecmiss, defined as the projection on the plane perpendicular to the beams of the negative vector sum of the particles momenta and its magnitude, denoted as \ETmiss.
To minimise the contributions from pileup interactions,
charged tracks are required to originate from the primary vertex (reconstructed using the deterministic annealing algorithm~\cite{Chatrchyan:2014fea}), which is the one characterised by the largest $\pt^2$ sum of its associated tracks.

Electrons are identified by combining information from tracks and ECAL clusters, including energy depositions from
final-state radiation~\cite{Khachatryan:2015hwa}.
Muons are identified through a combined fit to position measurements
from both the inner tracker and the muon detectors~\cite{CMS-PAPER-MUO-10-004}.
The $\tauh$ objects are identified and reconstructed using the ``hadron-plus-strips''
algorithm~\cite{HPS}, which uses charged hadrons and photons to reconstruct the main hadronic
decay modes of the $\tau$ lepton: one charged hadron, one charged hadron and photons, and three charged hadrons.
Electrons and muons can be misidentified as hadronic taus if produced in jets or if close-by activity
from pile-up or bremsstrahlung is present. These misidentifications are suppressed using dedicated criteria based on
the consistency between the measurements in the tracker, the calorimeters, and the muon detector~\cite{HPS}.
To reject nonprompt or misidentified leptons, requirements are imposed on the isolation criteria,
based on the sum of deposited energies. The absolute lepton isolation $I_\text{abs}$ is
defined by the scalar sum of the \pt of the charged particles from the primary vertex,
neutral hadrons, and photons in an isolation cone of size $\Delta R=\sqrt{\smash[b]{(\Delta\eta)^2+(\Delta\phi)^2}}=0.4$
($\Delta R=0.3$ for electrons), centred around the lepton direction. To reduce the effect from pileup, the energy
deposit released in the isolation cone by charged particles not associated with the primary vertex is subtracted from
the neutral particles \pt scalar sum. For electrons and muons the relative isolation, defined as $I_\text{rel} = I_\text{abs} /\pt$,
is used.

Jets are clustered using the anti-\kt algorithm~\cite{Cacciari:2008gp}, with a distance parameter
of 0.5, as implemented in the \textsc{Fastjet} software package~\cite{Cacciari:2011ma}. Charged particles not associated with the primary vertex
are excluded by means of the charged-hadron subtraction technique~\cite{CMS:JME-14-001}. The remaining energy originating from pileup interactions,
including the neutral components, is subtracted based on the median energy density in the detector computed through the effective jet area technique~\cite{Cacciari:2007fd}.
The identification of \PQb quark initiated jets is achieved through the combined secondary vertex (CSV)
algorithm~\cite{Chatrchyan:2012jua}, which exploits observables related to the long lifetime of B hadrons.

\subsection{Selection for \texorpdfstring{$\Z\to\ell\ell$}{Zll}}\label{subsec:Zll}

In selecting $\ell\ell \PQb\PQb$ and $\ell\ell\PGt\PGt$ events, the leptons from Z boson decay
are required to be well within the CMS trigger and detector
acceptance of $\pt>20$\GeV and $\abs{\eta}<2.5$ for electrons, and
$\pt>20$\GeV, $\abs{\eta}<2.4$ for muons.
Muon momentum-scale~\cite{CMS-PAPER-MUO-10-004} and electron energy corrections~\cite{Khachatryan:2015hwa} are applied
to recover the global shift of the scale observed between data and simulation.
The requirement on the relative isolation for the leptons is set to $I_\text{rel}<$ 0.15
for electrons and $I_\text{rel}<$ 0.2 for muons in selecting $\ell\ell\PQb\PQb$ events.
For the leptons from the Z boson, in the case of $\ell\ell\PGt\PGt$ events, the required relative isolation
is $I_\text{rel}<$ 0.3. The presence of two reconstructed same-flavour, oppositely charged
lepton candidates forming a pair with invariant mass in the range of 76--106\GeV is
required to suppress contamination of non-resonant Drell--Yan+jets
and \ttbar processes. In events where multiple Z candidates are present, the lepton pair with the invariant mass closest to the nominal
Z boson mass~\cite{Agashe:2014kda} is chosen.

\subsection{Event selection for \texorpdfstring{$\ell\ell\PQb\PQb$}{llbb}}\label{subsec:baselineLLBB}

For the $\ell\ell \PQb\PQb$ search, the jets are selected to be in the kinematic region $\pt>30$\GeV and $\abs{\eta}<2.4$.
At least two CSV b-tagged jets are required to be present in the
event, to reduce the contribution of Z+light-parton jets (originating
from gluons or u, d, or s quarks) events. The threshold on the b tagging discriminator corresponds
to a b tagging efficiency greater than 65\% and to
a misidentification probability for light-parton jets of $1\%$~\cite{Chatrchyan:2012jua}.
The two b-tagged jets with highest values of the CSV discriminant are considered as candidate decay products of the new light resonance.

The \MET significance~\cite{Khachatryan:2014gga,ETmissSig},
representing a $\chi^2$ difference between the observed result for
\MET and the \MET~=~0 hypothesis, is used to suppress background events
originating from \ttbar processes. This variable provides an event-by-event assessment of the likelihood that the observed missing transverse energy is consistent with zero given the reconstructed content of the event and known measurement resolutions.
This variable is a stronger discriminant against \ttbar background than \MET alone and also provides smaller systematic uncertainties. The distribution of the \ttbar component motivates the requirement on the \MET significance to be smaller than~10.

\subsection{Event selection for \texorpdfstring{$\ell\ell\tau\tau$}{lltautau}}\label{subsec:baselineLLTAUTAU}

To increase the signal sensitivity in the high $\tau\tau$ mass region, the $\ell\ell\PGt\PGt$ event selection includes
the requirement of a transversely boosted Z boson ($\pt>20\GeV$), together with a large ($>$1.5\unit{rad}) azimuthal angle
between the Z boson flight direction and \ptvecmiss, particularly effective in suppressing the Z+jets background.
In addition to the two light leptons required to reconstruct the Z boson, two additional oppositely charged and
different-flavor leptons (\Pe, \Pgm, and $\tauh$) are used to reconstruct the A or H boson candidate.
The requirements on the pseudorapidity for light leptons are the same as for the Z decay leptons, with the \pt threshold lowered
to 10\GeV. The $\tauh$ candidates are required to have $\pt>20\GeV$ and $\abs{\eta}<2.3$.
The relative isolation for electrons and muons, and the absolute isolation for \PGt leptons
are required to be smaller than 0.3 and 2\GeV, respectively.
Since the Z+jets background is characterised by a softer lepton transverse momentum spectrum than the signal one, this background is reduced by selecting events with high $L_\mathrm{T}$, where $L_{\mathrm{T}}$ indicates the scalar sum of the visible \pt of the decay products from a
$\PGt\PGt$ pair. Both the isolation requirements and the value of the $L_{\mathrm{T}}$ threshold are determined as a result of an optimisation procedure that maximises the expected significance of the searched signal. The optimal requirement on the $L_{\mathrm{T}}$ quantity is
found by scanning the threshold between 20 and 200\GeV, at intervals of 20\GeV.

Jets are required to have $\pt>30\GeV$ and $\abs{\eta}<4.7$. To reduce the large \ttbar background, all events with at least one
jet with $\pt>20\GeV$ and $\abs{\eta}<2.4$, reconstructed as a jet originating from a b quark according to the output of the CSV
discriminator used for tagging, are vetoed.

To calculate the $\tau\tau$ invariant mass, the secondary-vertex fit algorithm (\textsc{svfit})~\cite{Bianchini:2014vza} is used,
a likelihood-based method that combines the reconstructed \ptvecmiss and its resolution with
the momentum of the visible $\tau$ decay products to obtain an estimator of the mass of the parent particle.
\section{Modelling of the background}\label{sec:background}

\subsection{The \texorpdfstring{$\ell\ell \PQb\PQb$}{llbb} channel}

The relevant sources of background for the $\ell\ell \PQb\PQb$ final
state originate from Z+jets processes, \ttbar and tW production, diboson production, and vector boson production in
association with a SM Higgs boson. The contributions of Z+jets and \ttbar backgrounds are measured by means of a data-based method,
the diboson and tW backgrounds are normalised to the CMS measurements. For these backgrounds, the shapes are taken from MC,
while the normalisations are extracted from data. The vector boson production in association with a SM Higgs boson is normalised
to the theoretical prediction.

The comparison of data and predictions after the selection of events for the $\ell\ell \PQb\PQb$ final state shows the importance of an accurate theoretical calculation of the Z+jets production rate.
In particular, in the 400-700\GeV range of the $m_{\ell\ell\PQb\PQb}$ distribution, the data is found to exceed the LO prediction by up to two standard deviations, depending on the considered mass.
This excess is no longer significant when NLO QCD corrections, as implemented in a\MCATNLO~\cite{Alwall:2014hca}, are included in
the modelling of the Z+jets process. For this reason, the LO predictions are corrected using a reweighting technique, in order to account
for NLO QCD effects.
To this end, it becomes necessary to apply the reweighting according to the parton (or hadron) flavour of the jets in the generated event.
The ratio NLO/LO of the light- and heavy-flavour components of
the $m_{\ell\ell\mathrm{jj}}$ distribution is each fitted with a third-order polynomial and a separate
reweighting of the shape of the light and heavy flavour components of $m_{\ell\ell\mathrm{jj}}$ is applied,
resulting in better agreement with the data.

To determine the Z+jets and \ttbar normalization, a data-based method is exploited.
Data-derived correction factors for simulation are obtained after an additional categorisation of the Z+jet background events,
based on the flavour (b jet or not) and multiplicity (exactly two jets or three or more jets) of the reconstructed jets.
These categories are sensitive to NLO effects related to the modelling of extra jets~\cite{Chatrchyan:2014dha}.
Scale factors (SFs) are introduced for the \ttbar background and the light and heavy flavour components of Z+jets background. These are
left free to float in a two-dimensional fit of the distributions predicted by the simulation to the data.
The distributions used as input are the product of the CSV discriminants of the two selected jets, and the invariant mass of the
lepton pair from the Z boson decay in the range $60 <m_{\ell\ell}<120\GeV$. The first observable is sensitive to the contribution
from non-b jets, whereas the second one is sensitive to the contribution of the \ttbar production process.
The fit is performed simultaneously in four different categories: electrons, muons, exactly two jets, and more than two jets.
The SF for the \ttbar is found to be very close to the unity, while for the Z+jets process the SFs depart from unity by as much as 1.3 for
the light flavour component.

The overall yields from diboson and tW processes are normalised to the CMS
measurements~\cite{CMS:2014xja,Khachatryan:2015sga,Chatrchyan:2014aqa,CMS_PAPER_Wt}. The
associated production of a Z boson together with the Higgs-like scalar
boson (Zh) is also accounted for as background, and normalised to the
expected theoretical cross section~\cite{Chatrchyan:2013zna}.

\subsection{The \texorpdfstring{$\ell\ell \tau\tau$}{lltautau} channel}

Methods based on both data and simulation are used to
estimate the residual background after event selection.
Normalisations and mass distributions in the ZZ, Zh, as well as for the minor fully leptonic WWZ, WZZ, ZZZ and $\ttbar\Z$ backgrounds
are estimated from simulation. The Z+jets and WZ+jets contributions are measured by means of a data-based method.

Production of Z+jets and WZ+jets constitutes the main source
of background when at least one lepton is misidentified.
Misidentified light leptons arise from semileptonic decays of
heavy-flavour quarks, decays in flight of hadrons, and photon
conversions, while jets originating from quarks or gluons can be misidentified as $\tauh$.
Backgrounds with at least one misidentified lepton are estimated from control samples in data
starting from the estimation of the lepton misidentification probabilities.
The lepton misidentification probability is defined as the
probability that a genuine jet, satisfying loose lepton identification criteria (which refer to the so-called ``loose'' lepton), also passes
the identification criteria required for a lepton candidate in the signal region
(so-called ``tight'' lepton). This probability is measured for each
lepton flavour using a data sample where a Z candidate is selected, and
an additional single lepton (electron, muon, or $\tauh$) passes the loose identification requirements.
Counting the fraction of such loose leptons that also pass the tight lepton
identification criteria in the \pt bins of the reconstructed jet closest, in $\Delta R$, to the loose lepton,
yields the misidentification probability $f$ as a function of $\pt$.
The contribution from genuine leptons arising from the WZ and ZZ
production are subtracted. Once the misidentification probabilities are computed, three control regions ($CR$) are defined with a Z
candidate and two opposite-sign leptons, as follows: the $CR_{00}$ wherein both leptons pass loose identification criteria but
not the tight ones; $CR_{10}$ region, wherein one lepton passes tight
identification requirements, the other only loose criteria, and the loose lepton is the $\tauh$ with lower \pt in the
$\tauh\tauh$ channel, the light lepton in the $\ell\tauh$ channels,
and the electron in the $\Pe\mu$ channel; the $CR_{01}$ region, which is similar to $CR_{10}$ but the loose
lepton is the $\tauh$ with higher \pt in the $\tauh\tauh$ channel, the $\tauh$ in the
$\ell\tauh$ channels, and the muon in the $\Pe\mu$ channel.
The estimated $N_{\mathrm{misID}}$ of the background with at least one
misidentified lepton from a pair of closest-jet \pt bins is given by:
\begin{equation}
\label{eq:FRformula}
N_{\text{misID}} = N_{10}\frac{f_1}{1-f_1} + N_{01}\frac{f_2}{1-f_2} - N_{00}\frac{f_1f_2}{(1-f_1)(1-f_2)},
\end{equation}

where $N_{00}$, $N_{01}$, and $N_{10}$ denote the number of events from the $CR_{00}$, $CR_{01}$, and $CR_{10}$ control regions, respectively,
with closest jets in the considered \pt bins, and $f_1$ and $f_2$ indicate the misidentification
probabilities associated with the two different flavor (except for the $\tauh\tauh$ final state) loose leptons in the \pt bins.
The expression in Eq.~(\ref{eq:FRformula}) takes into account both the background with two
misidentified leptons (mostly from Z+jets) and that from only one
misidentified lepton (primarily from WZ+jets).

The contamination from genuine leptons in the control regions from the SM Zh, WWZ, WZZ, ZZZ, $\ttbar\Z$,
and ZZ processes is estimated from simulation, and subtracted from
$N_{00}$, $N_{01}$, and $N_{10}$. The total background in the signal region is obtained by summing the contributions
from all pairs of \pt bins.
\section{Systematic uncertainties}\label{sec:systematics}

The systematic uncertainties are reported in the following paragraphs and summarised in Tab.~\ref{tab:syst_summary}.

The uncertainty on the integrated luminosity recorded by CMS is estimated to be
2.6\%~\cite{CMSLumi}.

The systematic uncertainties associated with the lepton efficiency SFs, used to correct the simulation and derived from
studies at the Z peak using the tag-and-probe (T\&P) method~\cite{CMS-PAPER-MUO-10-004,Khachatryan:2015hwa}, are approximately
1\% for muons and 2\% for electrons, and affect both signal and background processes in the same way.
Also, the uncertainties on the double muon and double electron trigger efficiencies are evaluated to be 1\% from similar studies at
the Z peak~\cite{Chatrchyan:2014mua}.

The uncertainty on the jet energy scale is derived from the method of Ref.~\cite{Chatrchyan:2011ds} and the parameters describing the shape of the energy distribution are varied by one standard deviation (SD).
The effect is estimated separately on the background and on the signal, resulting in a 3--5\% variation, depending on the \pt and $\eta$ of the jets.
The uncertainty on signal and background yields induced by the
imperfect knowledge of the jet energy resolution is estimated to be 3\%~\cite{Chatrchyan:2011ds}.

The uncertainties affecting b tagging efficiencies are
$\pt$-dependent, and vary from 3\% to 12\% (for $\pt > 30\GeV$)~\cite{Chatrchyan:2012jua}. The impact of
these uncertainties on the normalisation of signal is 5\% for background and 4--6\% for signal in the
$\ell\ell\PQb\PQb$ analysis, and about 1\% in the
$\ell\ell\PGt\PGt$ analysis. The uncertainty in the mistagging rate is found
to have a negligible impact.

The systematic uncertainty on the signal is evaluated by
varying the set of parton distribution functions (PDFs) according to the PDF4LHC prescriptions~\cite{Botje:2011sn,Alekhin:2011sk,Ball:2012cx}
and the factorisation and renormalisation scales by varying their values by a factor one half and two. An effect of 5--6\% is
estimated for the entire mass range for both $\ell\ell\tau\tau$ and
$\ell\ell\PQb\PQb$ final states. This uncertainty is estimated by
propagating these variations through the signal simulation and reconstruction sequence and thus accounts for uncertainties related
to both signal cross section and acceptance.

Finally, an 11\% uncertainty is assigned to the ZZ normalisation from the cross section measured by CMS~\cite{CMS:2014xja}.

\begin{table*}[htb]
\topcaption{Summary of systematic uncertainties for both $\ell\ell\tau\tau$ and $\ell\ell\PQb\PQb$ final states.}
\label{tab:syst_summary}
\centering
\begin{tabular}{lcc}
\hline
Source                                         & \multicolumn{2}{c}{Uncertainty [\%]} \\\cline{2-3}
& $\PH\to\Z\PSA\to\ell\ell\PQb\PQb$ & $\PH\to \Z\PSA\to\ell\ell\tau\tau$ \\[2ex]
 Luminosity                                     & 2.6              & 2.6              \\
 Lepton identification/isolation/scale          & 1--2             & 1--2             \\
 Lepton trigger efficiency                      & 1                & 1                \\
 Jet energy scale                               & 3--5             & 3--5             \\
 Jet energy resolution                          & 3                & 3                \\
 b-tagging and mistag efficiency               & 4--6                & 1             \\
 Signal modelling (PDF, scale)                  & 5--6             & 5--6             \\
 Background norm. ($ZZ$)                        & 11               & 11               \\
 Background norm. (Z+jets and \ttbar)           & $<$8             & ---               \\
 Background norm. (tW, WW, WZ and Zh)             & 8--23            & ---               \\
 Z+jet background modelling                     & 5--30            & ---               \\
 Signal efficiency extrapolation                & 3--50            & ---               \\
 Tau identification/isolation                   & ---               & 6                \\
 Tau energy scale                               & ---               & 3                \\
 Reducible background estimate                  & ---               & 40               \\
\hline
\end{tabular}
\end{table*}
For the $\ell\ell\PQb\PQb$ final state, the uncertainty on the SFs used for normalisation of Z+jets
and \ttbar backgrounds is derived from the statistical uncertainty resulting from the fit used to
derive these SFs and it is estimated to be $<$8\%.
An additional systematic uncertainty associated with the $m_{\ell\ell \PQb\PQb}$ spectrum correction, described in Sec.~\ref{sec:background},
ranges from 5\% for $m_{\ell\ell\PQb\PQb}$ below 700\GeV to 30\% for masses at the TeV scale.
An uncertainty of 8\% is assigned to the normalisation of the WW process, corresponding to the uncertainties in the cross
section measured by CMS~\cite{Khachatryan:2015sga}. A similar uncertainty is assigned also to the WZ process, which shares the
same sources of uncertainties in the cross section measurement.
For the minor tW background, the uncertainty is estimated as 23\%, also based on the
measured cross section~\cite{CMS_PAPER_Wt}. A 7\% uncertainty is
assigned to the Zh process, reflecting the uncertainty on the theoretical cross
section~\cite{Chatrchyan:2013zna}. Given the small cross section for
this SM process compared to other background processes, its contribution to the
background normalisation uncertainty has been calculated to be less
than 1\% and is thus considered negligible.
In order to interpolate smoothly the signal efficiency across the parameter space,
additional mass points for the $\ell\ell\PQb\PQb$ final state are processed using
a parametric simulation~\cite{deFavereau:2013fsa}, tuned for delivering a realistic
approximation of the CMS response in the reconstruction of physics objects used in this
search. For this reason, an additional source of uncertainty is introduced for the SF applied to
these samples to reproduce the efficiency measured with the full simulation.
This is measured for the different signal points in the $m_{\PH}$-$m_{\PSA}$ plane and
it is close to 3\% in most of the phase space, but rises to 50\% at the boundaries of the sensitivity region.

In the $\ell\ell\PGt\PGt$ final state, the uncertainty of 6\%~\cite{HPS} in the $\tauh$ identification efficiency,
which has been determined using a T\&P method, has been taken into account.
The $\tauh$ energy scale uncertainty is within 3\%~\cite{HPS} and only affects the shapes of the $\PGt\PGt$ mass distributions.
The systematic uncertainties estimated for $\Pe$, $\mu$, $\tauh$ and jet energy scales are propagated to \ptvecmiss and to the mass distributions.
The propagation to \ptvecmiss involves a sum of the energies of each object first and a consequent subtraction of such contributions
once the nominal energy scales (or resolutions) are varied up and down by one SD (for $\Pe$, $\mu$, $\tauh$, and jets).
One of the main systematic uncertainties is related to the nonprompt background estimation.
This uncertainty has been evaluated using simulation by comparing the direct estimate of the backgrounds with that
obtained using the procedure adopted in the analysis, but applied to simulated events.
The discrepancy between the two estimates never exceeds 40\%. This value is thus considered as the
uncertainty on the estimates of the reducible background yield for all channels and all $L_{\mathrm{T}}$ thresholds.
\section{Results}\label{sec:results}

The analysis searches for new resonance decays by comparing data to simulation in the two-dimensional
plane defined by the four-body ($m_{\ell\ell \PQb\PQb}$ or $m_{\ell\ell\tau\tau}$) and two-body ($m_{\PQb\PQb}$ or $m_{\tau\tau}$)
invariant masses. The numerical values for the upper limits or the significance of a local excess are obtained using the
asymptotic method described in Ref.~\cite{Cowan:2010js}.
The $\mathrm{CL}_\mathrm{S}$ method~\cite{Junk:1999kv,Read:2002hq} is used to determine the 95\% confidence level (CL) upper limits on the excluded
signal cross section. For both final states, the limits in the lower-right triangle of
the mass plane, which corresponds to the process $\PSA\to\Z\PH$, are obtained by mirroring
the results obtained in the upper-left triangle, since the signal efficiencies for $\PH\to\Z\PSA$ and
$\PSA\to\Z\PH$ are equal for the same masses of the heavy and light Higgs bosons in the two processes.

\subsection{The \texorpdfstring{$\ell\ell \PQb\PQb$}{llbb} channel}

For the $\ell\ell\PQb\PQb$ final state, results are obtained using
a counting approach, which can be reinterpreted in other theoretical models with the same final
state. Results are reported in bins of $m_{\PQb\PQb}$ and $m_{\ell\ell\PQb\PQb}$ masses, in the range
from 10\GeV to 1\TeV for $m_{\PQb\PQb}$, and from 140\GeV to 1\TeV for $m_{\ell\ell\PQb\PQb}$.
To define the proper granularity of the binning, a study is performed using signal benchmark points and evaluating
the width of the $m_{\PQb\PQb}$ and $m_{\ell\ell\PQb\PQb}$ peaks in the considered mass range. The average
reconstructed width, defined as one SD, for $m_{\PQb\PQb}$ and $m_{\ell\ell\PQb\PQb}$ is found to be approximately
15\% of the considered mass. The bin widths have been chosen to be $\pm$1.5\,SD around each considered mass point.

The efficiency, defined as the fraction of generated signal events reconstructed after the final selection, is calculated with the full CMS simulation
and reconstruction software at 13 representative signal points in the $m_{\PH}$-$m_{\PSA}$ mass plane.
The signal efficiencies for the rest of the plane are obtained by interpolating the ratio between the full simulation and the parametric simulation
(typically 0.9), calculated in each of the 13 signal mass points, and scaling the efficiencies calculated using the parametric simulation
by this interpolated ratio. The resulting signal efficiency ranges from 8\% at ($m_{\PSA}$, $m_{\PH}$) = (100, 300)\GeV to 13\% at
(300, 600)\GeV.

\begin{figure}[hbtp]
\centering
   \includegraphics[width=0.48\textwidth]{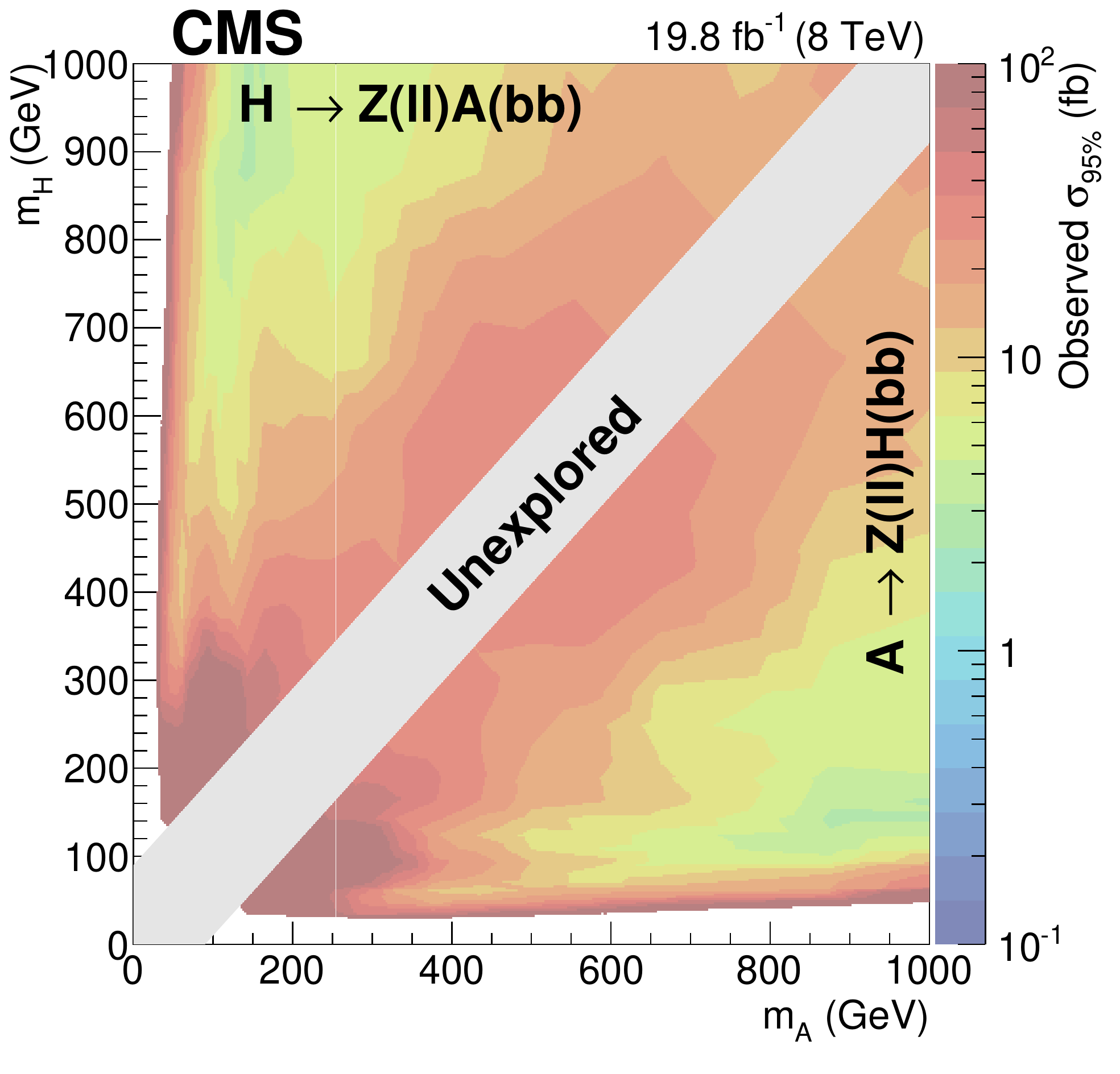}
   \caption{Observed 95\% CL upper limits on $\sigma_{\mathrm{\PH/\PSA} \to \Z\mathrm{\PSA/\PH} \to \ell\ell \PQb\PQb}$ as a function of $m_{\PSA}$ and $m_{\PH}$.}
   \label{fig:limXSllbb}
\end{figure}

Figure~\ref{fig:limXSllbb} shows the observed upper limits
on the product of the cross section ($\sigma$) and branching fraction ($\mathcal{B}$) for the $\ell\ell\PQb\PQb$ final state
in the $m_{\PH}$-$m_{\PSA}$ plane.
The achieved sensitivity provides an exclusion limit at 95\%~CL of approximately 10\unit{fb} for a large fraction of the two-dimensional mass plane.
In particular, the observed limit ranges from just above 1\unit{fb} for $m_{\PH}$
close to 1\TeV to 100\unit{fb} for $m_{\PH}<$ 300\GeV. The
validity of these results is applicable to models allowing the existence of both A and H
bosons with a natural width smaller than 15\% of their masses.

\begin{figure}[hbtp]
\centering
   \includegraphics[width=0.46\textwidth]{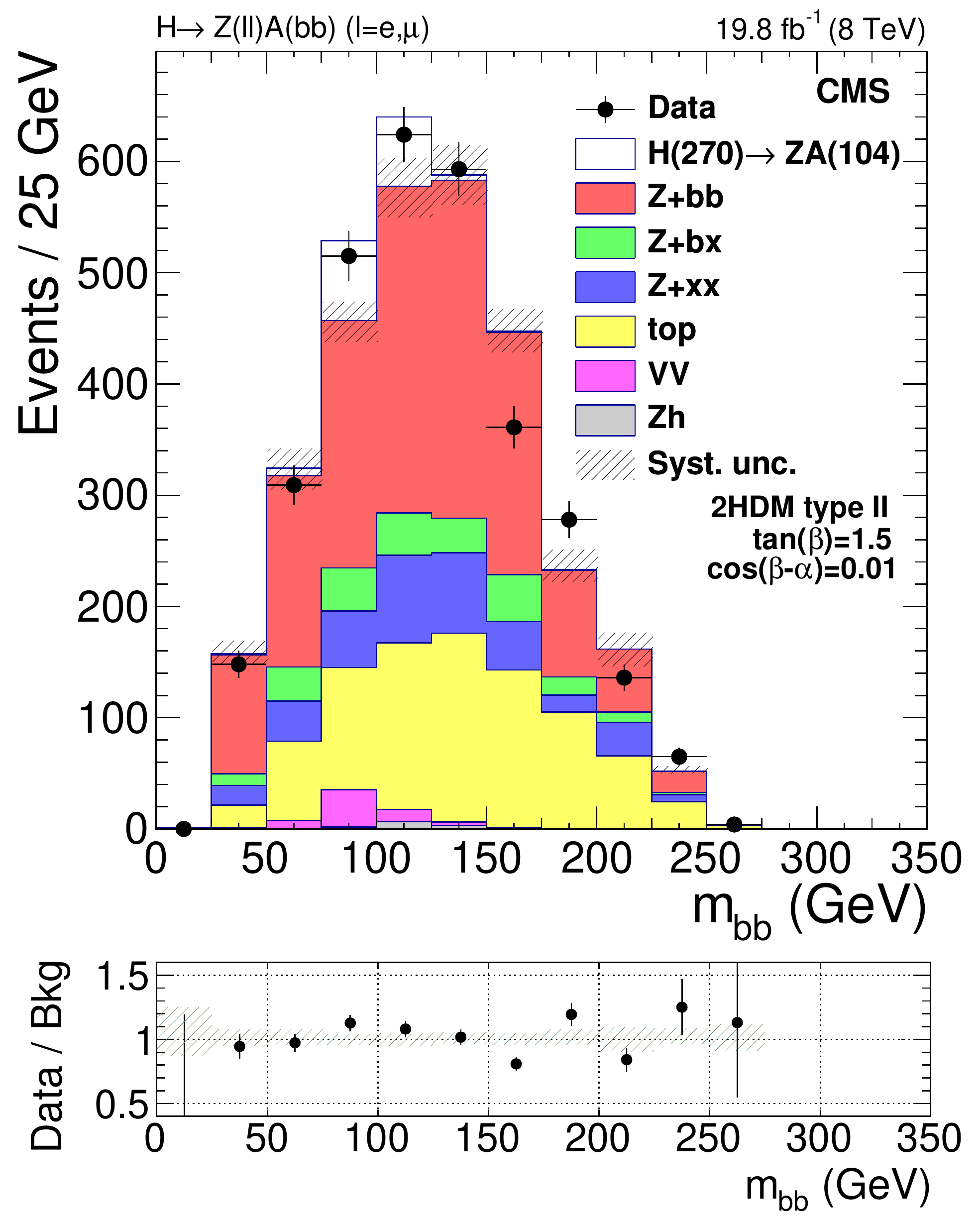}
   \includegraphics[width=0.46\textwidth]{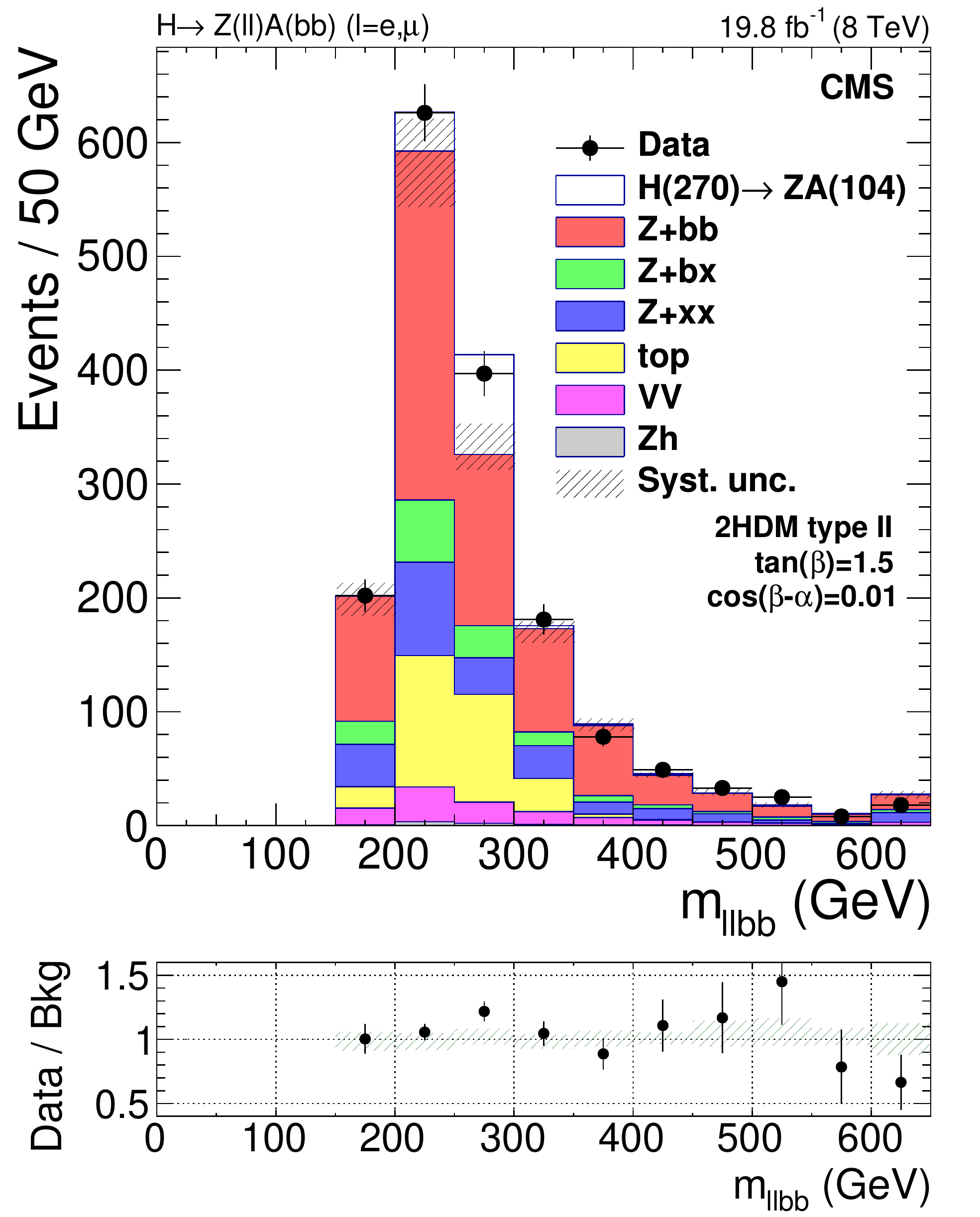}
    \caption{(\cmsLeft) The $m_{\PQb\PQb}$ spectrum for events selected in
      the $222< m_{\ell\ell \PQb\PQb}<350\GeV$ region for data and simulation and the relative ratio.
      (\cmsRight) The $m_{\ell\ell \PQb\PQb}$ spectrum for events selected
      inside the region $72< m_{\PQb\PQb} < 114\GeV$ region for data and simulation and
      the relative ratio. The signal corresponding to $m_{\PH}=270\GeV$ and $m_{\PSA}=104\GeV$, normalized to the NNLO \textsc{SusHi} cross section, is superimposed
      for $\tan\beta=1.5$ and $\cos(\beta-\alpha)=0.01$ in the 2HDM type-II scenario. The overall systematic uncertainties in the simulation are
      reported as a hatched band.}
   \label{fig:Mbb}
\end{figure}

Two moderate excesses are observed for the
$\ell\ell\PQb\PQb$ channel in the regions around
($m_{\PQb\PQb}$, $m_{\ell\ell\PQb\PQb}$) = (95, 285)\GeV and (575, 660)\GeV.
According to the procedure described at Ref.~\cite{ATLAS:2011tau}, they have local significances of 2.6 and 2.85\,SD respectively,
which become globally 1.6 and 1.9\,SD, once accounting for the
look-elsewhere effect~\cite{Gross:2010qma}.
The low-mass excess is more compatible with the signal hypothesis, both in terms
of yield and width. The reconstructed invariant mass distributions for the $\PQb\PQb$
and $\ell\ell\PQb\PQb$ systems, in the regions around this excess, are reported in
Fig.~\ref{fig:Mbb} and compared with the expectations from background
processes.
A 2HDM type-II benchmark signal at $m_{\PH}=270\GeV$ and
$m_{\PSA}=104\GeV$, normalised to the NNLO \textsc{SusHi} prediction, is also superimposed.
\subsection{The \texorpdfstring{$\ell\ell \tau\tau$}{lltautau} channel}

\begin{figure*}
\centering
    \includegraphics[width=0.62\textwidth]{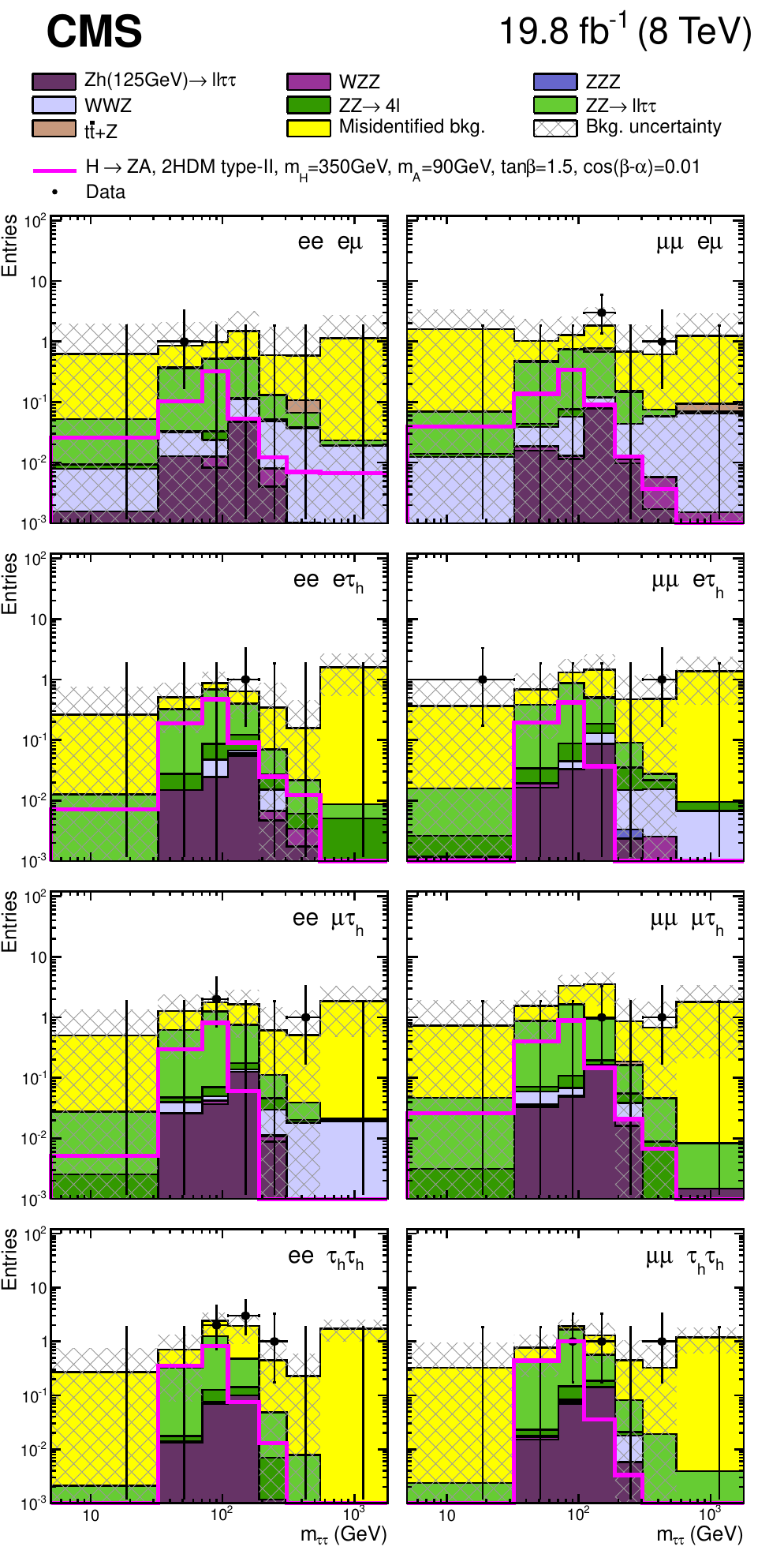}
    \caption{\textsc{svfit} mass distributions for different final states of the $\PH\to\Z\PSA\to\ell\ell\tau\tau$
    process, where the $\Z$ boson decays to $\Pe\Pe$ (right column) and $\mu\mu$ (left column).
    The expected signal corresponding to $m_{\PH}=350\GeV$ and $m_{\PSA}=90\GeV$, whose cross section times
    branching fraction is normalised to the NNLO \textsc{SusHi} prediction, is superimposed
    for $\tan\beta=1.5$ and $\cos(\beta-\alpha)=0.01$ in the 2HDM type-II scenario.
    Only statistical uncertainties are reported as a hatched band.}
    \label{fig:lltautauMassPlots}
\end{figure*}

In the context of the $\ell\ell\tau\tau$ analysis, a search based on the $m_{\tau\tau}$
distribution is performed. For every considered pair of $\PH$ and $\PSA$ mass values, the search is
performed in eight $\tau\tau$ \textsc{svfit} binned mass distributions, each corresponding to one of the eight considered final states. Variable bin widths are adopted in order to account for the mass resolution. A simultaneous likelihood fit to the observed distributions is performed with
the expected distributions from the background-only and signal plus background hypotheses. The normalisation of the signal
distribution is a free parameter in the fit. No significant deviations are observed in data from the SM expectation.
The \textsc{svfit} mass distributions of the $\tau\tau$ pair in the eight different final states are shown in
Fig.~\ref{fig:lltautauMassPlots}. The chosen signal corresponds to $m_{\PH}=350\GeV$ and $m_{\PSA}=90\GeV$,
which is the one closest to the centre of the bin in which the highest excess is observed in the $\ell\ell\cPqb\cPqb$ channel.
The shown shapes correspond to $L_T>40\GeV$ for $\Pe\mu$, $L_T>60\GeV$ for $\Pe\tauh$ and $\mu\tauh$, and $L_T>80\GeV$ for $\tauh\tauh$.

\begin{figure}[hbtp]
\centering
   \includegraphics[width=0.49\textwidth]{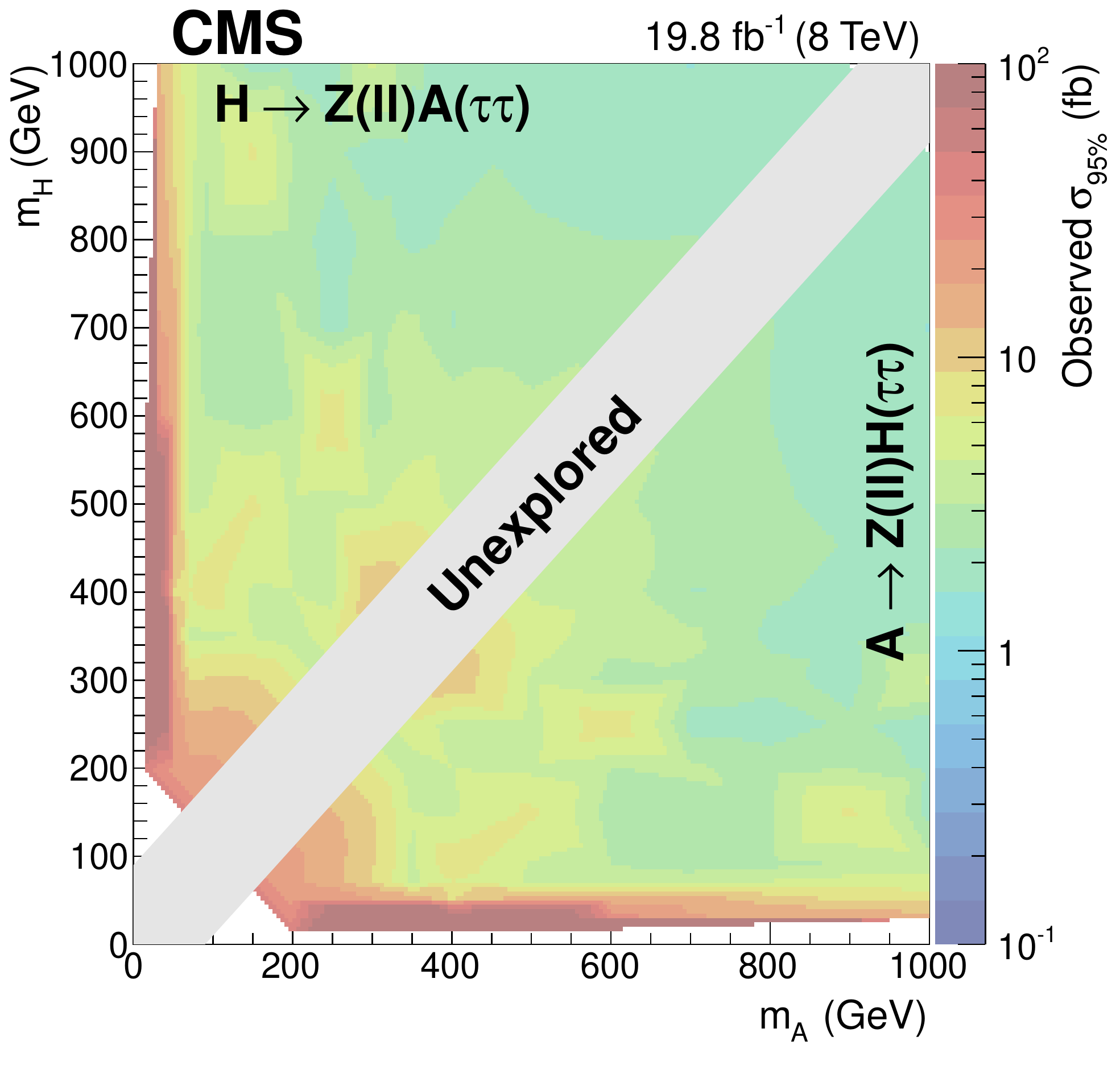}
   \caption{Observed 95\% CL upper limits on $\sigma_{\mathrm{\PH/\PSA}\to \mathrm{\Z\PSA/\PH} \to \ell\ell\tau\tau}$ as a function of $m_{\PSA}$ and $m_{\PH}$.}
   \label{fig:limXSlltautau}
\end{figure}

Figure~\ref{fig:limXSlltautau} shows the limit on $\sigma\,\mathcal{B}$ for the $\ell\ell\tau\tau$ final state in the $m_{\PH}$-$m_{\PSA}$ plane.
Signal cross sections of about 5--10\unit{fb} are excluded in most of the $m_{\PH}$-$m_{\PSA}$ plane ($500<m_{\PH/\PSA}<1000\GeV$ and $90<m_\mathrm{\PSA/\PH}<400\GeV$).

\subsection{Combination in the context of 2HDM}

\begin{figure}[!htb]
\centering
      \includegraphics[width=0.49\textwidth]{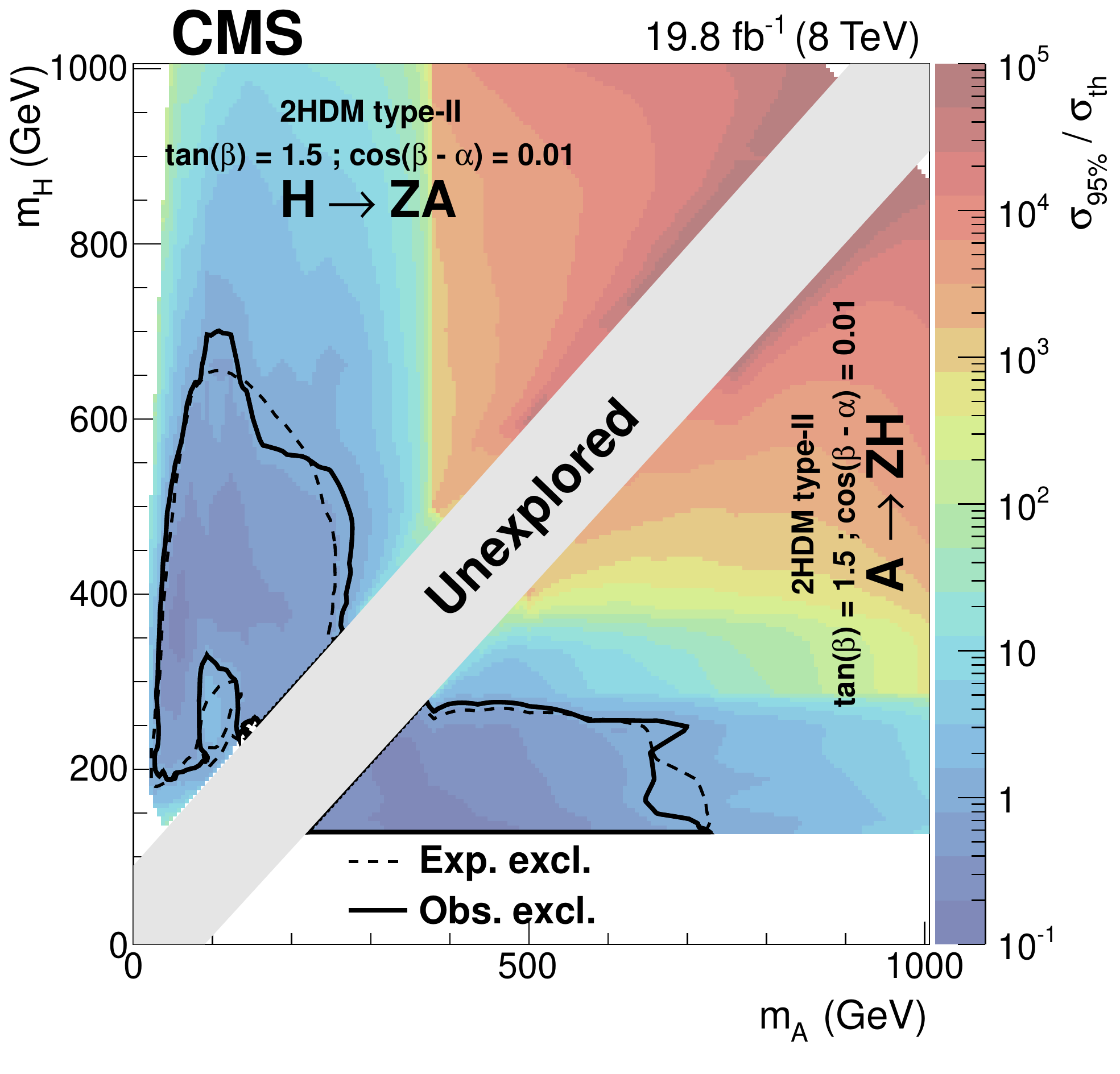}
    \caption{ Observed limits on the signal strength $\mu=\sigma_{95\%}/\sigma_{\mathrm{th}}$
     for the 2HDM benchmark, after combining results from $\ell\ell\PQb\PQb$ and $\ell\ell\PGt\PGt$ final states.
     The cross sections are normalised to the NNLO \textsc{SusHi} prediction, for a 2HDM type-II scenario with $\tan\beta=1.5$ and $\cos(\beta-\alpha)=0.01$.
     The dashed contour shows the region expected to be excluded. The solid contour shows the region excluded by the data.
   }
    \label{fig:limMUllbb}
\end{figure}

Observed and expected upper limits on the signal cross section
modifier $\mu=\sigma_{95\%}/\sigma_{\mathrm{th}}$ are also derived and reported
in Fig.~\ref{fig:limMUllbb}, where $\sigma_{\mathrm{th}}$ is the theory cross
section of the 2HDM signal benchmark used in this analysis.
The results are obtained from the combination of the $\ell\ell\PQb\PQb$ and
$\ell\ell\PGt\PGt$ final states. This search is not able to exclude
the high-mass regions where $m_{\PSA}>300\GeV$ and $m_{\PH}>300\GeV$,
due to the drop in the signal cross section, where the $\mathrm{\PSA/\PH}\to\ttbar$ channel opens up for $ m_\mathrm{\PSA/\PH}> 2 m_{\PQt}$, where $m_{\PQt}$ is the top quark mass~\cite{Grinstein:2013npa}. Furthermore, in the region
where highly-boosted topologies start contributing ($m_{\PH}\approx 10\, m_{\PSA}$), the
sensitivity is lower relative to the rest of the plane, primarily
caused by the inefficiency in reconstructing signal decay
products in such a regime. Still, a significant portion of the benchmark
masses is excluded for a 2HDM type-II scenario with $\tan\beta=1.5$ and $\cos(\beta-\alpha)=0.01$,
delimited by the solid contour in Fig.~\ref{fig:limMUllbb}. The observed 95\% CL exclusion region is localised
in the range $m_{\PH}$~=~200--700\GeV and $m_{\PSA}=20$--270\GeV
for the decay $\PH\to\Z\PSA$, and similarly in the range
$m_{\PSA}=200$--700\GeV and $m_{\PH}=120$--270\GeV for the
$\PSA\to\Z\PH$ decay.
The feature observed in the exclusion limit for the region around
$(m_{\PSA},m_{\PH})= ($75--100, 200--300)\GeV is the
result of an interplay between the larger \Z+jets background yields
expected in this region and the quickly evolving signal
cross section. The effect is visible in the expected limits
and becomes slightly broader in the observed ones given the
concurrent presence in the same region of a moderate data excess.
The region where $\abs{m_{\PH} - m_{\PSA}}<m_{\Z}$ is kinematically inaccessible.

\begin{figure}[hbtp]
\centering
    \includegraphics[width=0.49\textwidth]{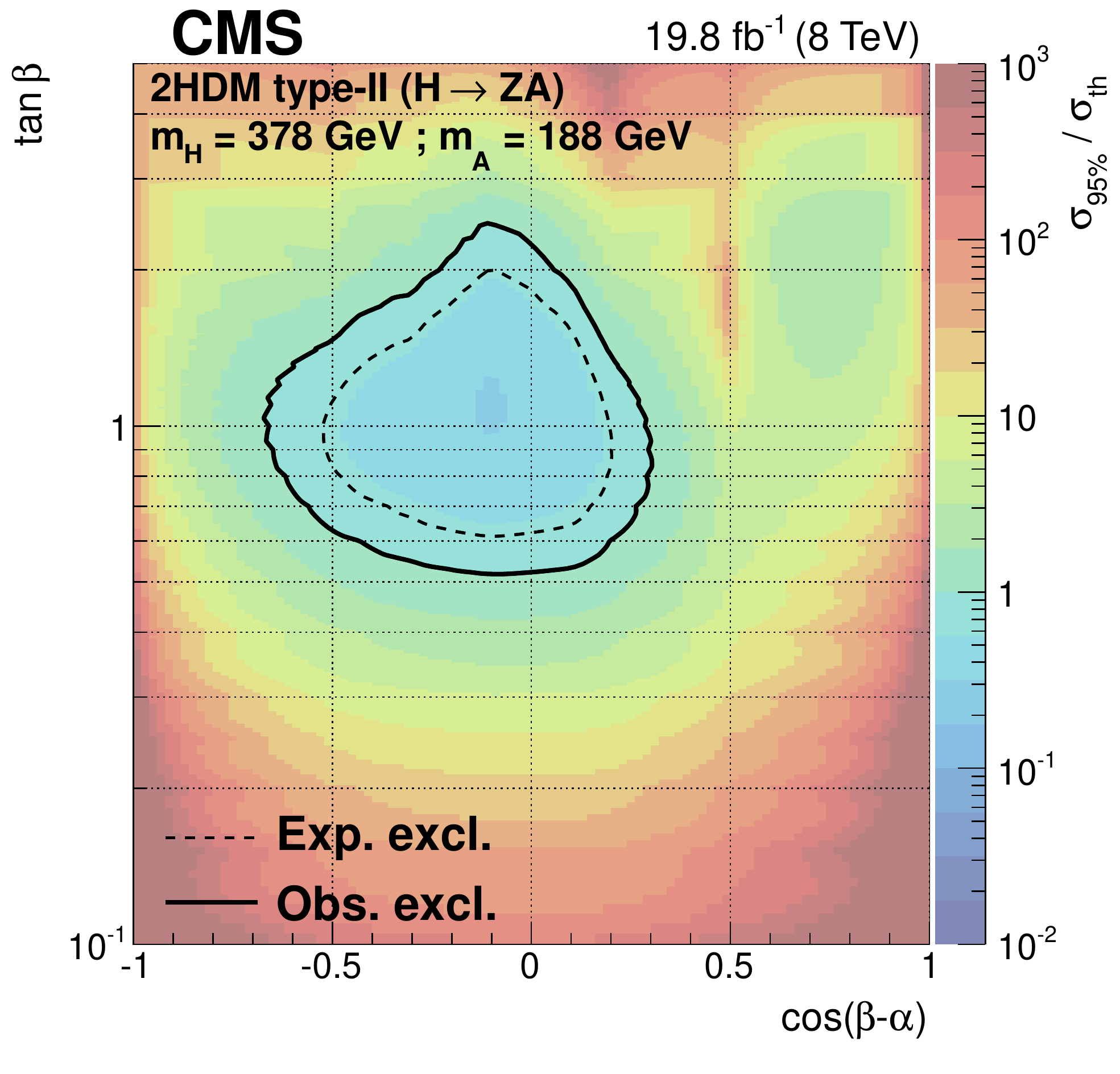}
    \caption{ Observed limits on the signal strength $\mu=\sigma_{95\%}/\sigma_{\mathrm{th}}$
     for the 2HDM benchmark after combining results from $\ell\ell\PQb\PQb$ and $\ell\ell\PGt\PGt$
     final states. The cross sections are normalised to the NNLO \textsc{SusHi} prediction.
     Limits are shown in the 2HDM parameters $\cos(\beta-\alpha)$ and \tanb for the signal masses of
     $m_{\PH}=378$\GeV and $m_{\PSA}=188$\GeV. The dashed contour shows the region expected to be excluded.
     The solid contour shows the region excluded by the data.
   }
    \label{fig:limCosTan}
\end{figure}

The limits on $\mu$ can be also visualised as a function of the 2HDM parameters \tanb and $\cos(\beta-\alpha)$ for a given pair of $m_{\PSA}$ and $m_{\PH}$, from the combination of $\ell\ell\PQb\PQb$ and $\ell\ell\PGt\PGt$ final states. Results are given in Fig.~\ref{fig:limCosTan}, where the exclusion limits on the parameters are shown for $m_{\PH}=378\GeV$ and $m_{\PSA}=188\GeV$, a mass pair chosen to be within the exclusion region of Fig.~\ref{fig:limMUllbb}.
The area contained within the solid line shows the parameter space excluded for the chosen mass pair, where \tanb lies between 0.5 and 2.3 and $\cos(\beta-\alpha)$ between $-0.7$ and 0.3.
\section{Summary}\label{sec:summary}

The paper describes the first CMS search for a new resonance decaying into a lighter resonance and a Z boson.
Two searches have been performed, targeting the decay of the lighter resonance into either a pair of oppositely charged $\tau$ leptons
or a \bbbar pair. The Z boson is identified via its decays to electrons or muons.
The search is based on data corresponding to an integrated luminosity of 19.8\fbinv in proton--proton collisions at $\sqrt{s}=8$\TeV.
Deviations from the SM expectations are observed with a global significance of less than 2 SD and upper limits on the product of
cross section and branching fraction are set.
The search excludes $\sigma \,\mathcal{B}$ as low as 5\unit{fb} and 1\unit{fb} for the $\ell\ell\PQb\PQb$
and $\ell\ell\PGt\PGt$ final states, respectively, depending on the light and heavy resonance mass values.

Limits are also set on the mass parameters of the type-II 2HDM model that predicts the processes $\PH\to\Z\PSA$ and $\PSA\to\Z\PH$, where H and A are CP-even and CP-odd scalar bosons, respectively. Combining the $\ell\ell\PQb\PQb$ and $\ell\ell\PGt\PGt$ final states, the specific model corresponding to the parameter choice $\cos(\beta -\alpha)=0.01$ and $\tanb=1.5$ is excluded for $m_{\PH}$ in the range 200--700\GeV and $m_{\PSA}$ in the range 20--270\GeV with $m_{\PH}>m_{\PSA}$, or alternatively for $m_{\PSA}$ in the range 200--700\GeV and $m_{\PH}$ in the range 120--270\GeV with $m_{\PSA}>m_{\PH}$.

Limits on the signal cross section modifier are also derived as a function of \tanb and $\cos(\beta-\alpha)$ parameters.
As a result, for specific $m_{\PH}$-$m_{\PSA}$ mass values, a fairly large region in the parameter space \tanb \vs $\cos(\beta-\alpha)$
is excluded. This covers a region unexplored so far, that cannot be probed by studying production and decay modes of the SM-like Higgs boson.
In particular, for $m_{\PH}=378\GeV$ and $m_{\PSA}=188\GeV$, a range where \tanb lies between 0.5 and 2.3 and $\cos(\beta-\alpha)$ between $-0.7$ and 0.3 is excluded, after the combination of the $\ell\ell\PQb\PQb$ and $\ell\ell\PGt\PGt$ final states.

\begin{acknowledgments}
We congratulate our colleagues in the CERN accelerator departments for the excellent performance of the LHC and thank the technical and administrative staffs at CERN and at other CMS institutes for their contributions to the success of the CMS effort. In addition, we gratefully acknowledge the computing centres and personnel of the Worldwide LHC Computing Grid for delivering so effectively the computing infrastructure essential to our analyses. Finally, we acknowledge the enduring support for the construction and operation of the LHC and the CMS detector provided by the following funding agencies: BMWFW and FWF (Austria); FNRS and FWO (Belgium); CNPq, CAPES, FAPERJ, and FAPESP (Brazil); MES (Bulgaria); CERN; CAS, MoST, and NSFC (China); COLCIENCIAS (Colombia); MSES and CSF (Croatia); RPF (Cyprus); MoER, ERC IUT and ERDF (Estonia); Academy of Finland, MEC, and HIP (Finland); CEA and CNRS/IN2P3 (France); BMBF, DFG, and HGF (Germany); GSRT (Greece); OTKA and NIH (Hungary); DAE and DST (India); IPM (Iran); SFI (Ireland); INFN (Italy); MSIP and NRF (Republic of Korea); LAS (Lithuania); MOE and UM (Malaysia); CINVESTAV, CONACYT, SEP, and UASLP-FAI (Mexico); MBIE (New Zealand); PAEC (Pakistan); MSHE and NSC (Poland); FCT (Portugal); JINR (Dubna); MON, RosAtom, RAS and RFBR (Russia); MESTD (Serbia); SEIDI and CPAN (Spain); Swiss Funding Agencies (Switzerland); MST (Taipei); ThEPCenter, IPST, STAR and NSTDA (Thailand); TUBITAK and TAEK (Turkey); NASU and SFFR (Ukraine); STFC (United Kingdom); DOE and NSF (USA).

Individuals have received support from the Marie-Curie programme and the European Research Council and EPLANET (European Union); the Leventis Foundation; the A. P. Sloan Foundation; the Alexander von Humboldt Foundation; the Belgian Federal Science Policy Office; the Fonds pour la Formation \`a la Recherche dans l'Industrie et dans l'Agriculture (FRIA-Belgium); the Agentschap voor Innovatie door Wetenschap en Technologie (IWT-Belgium); the Ministry of Education, Youth and Sports (MEYS) of the Czech Republic; the Council of Science and Industrial Research, India; the HOMING PLUS programme of the Foundation for Polish Science, cofinanced from European Union, Regional Development Fund; the OPUS programme of the National Science Center (Poland); the Compagnia di San Paolo (Torino); MIUR project 20108T4XTM (Italy); the Thalis and Aristeia programmes cofinanced by EU-ESF and the Greek NSRF; the National Priorities Research Program by Qatar National Research Fund; the Rachadapisek Sompot Fund for Postdoctoral Fellowship, Chulalongkorn University (Thailand); the Chulalongkorn Academic into Its 2nd Century Project Advancement Project (Thailand); and the Welch Foundation, contract C-1845.
\end{acknowledgments}

\bibliography{auto_generated}   

\cleardoublepage \appendix\section{The CMS Collaboration \label{app:collab}}\begin{sloppypar}\hyphenpenalty=5000\widowpenalty=500\clubpenalty=5000\textbf{Yerevan Physics Institute,  Yerevan,  Armenia}\\*[0pt]
V.~Khachatryan, A.M.~Sirunyan, A.~Tumasyan
\vskip\cmsinstskip
\textbf{Institut f\"{u}r Hochenergiephysik der OeAW,  Wien,  Austria}\\*[0pt]
W.~Adam, E.~Asilar, T.~Bergauer, J.~Brandstetter, E.~Brondolin, M.~Dragicevic, J.~Er\"{o}, M.~Flechl, M.~Friedl, R.~Fr\"{u}hwirth\cmsAuthorMark{1}, V.M.~Ghete, C.~Hartl, N.~H\"{o}rmann, J.~Hrubec, M.~Jeitler\cmsAuthorMark{1}, V.~Kn\"{u}nz, A.~K\"{o}nig, M.~Krammer\cmsAuthorMark{1}, I.~Kr\"{a}tschmer, D.~Liko, T.~Matsushita, I.~Mikulec, D.~Rabady\cmsAuthorMark{2}, B.~Rahbaran, H.~Rohringer, J.~Schieck\cmsAuthorMark{1}, R.~Sch\"{o}fbeck, J.~Strauss, W.~Treberer-Treberspurg, W.~Waltenberger, C.-E.~Wulz\cmsAuthorMark{1}
% ...
\vskip\cmsinstskip
% ...
1:~~Also at Vienna University of Technology, Vienna, Austria\\
2:~~Also at CERN, European Organization for Nuclear Research, Geneva, Switzerland\\
% ...

\end{sloppypar}
\end{document}